\documentclass[aps,prl,reprint,superscriptaddress,showpacs,floatfix]{revtex4-1}

\usepackage{mathrsfs}
\usepackage{bm}
\usepackage{graphicx}

\begin{document}

\title{Relaxation of Optically Excited Carriers in Graphene}

\date{\today}

\author{Raseong Kim}
\affiliation{Network for Computational Nanotechnology, Purdue University, West Lafayette, Indiana 47907}

\author{Vasili Perebeinos}
\affiliation{IBM Thomas J. Watson Research Center, Yorktown Heights, New York 10598}

\author{Phaedon Avouris}
\affiliation{IBM Thomas J. Watson Research Center, Yorktown Heights, New York 10598}

\begin{abstract}
We explore the relaxation of photo-excited graphene by solving a transient Boltzmann transport equation with electron-phonon (e-ph) and electron-electron (e-e) scattering. Simulations show that when the excited carriers are relaxed by e-ph scattering only, a population inversion can be achieved at energies determined by the photon energy. However, e-e scattering quickly thermalizes the carrier energy distributions washing out the negative optical conductivity peaks. The relaxation rates and carrier multiplication effects are presented as a function of photon energy and dielectric constant.
\end{abstract}

%\keywords{}

\maketitle

Following the success in fabricating field-effect transistors using a single carbon atom layer \cite{Novoselov_2004}, graphene has been actively studied as a possible basis for electronic and optoelectronic devices \cite{Avouris_2010}. Graphene's linear dispersion and zero energy gap result in unique transport and optical properties \cite{Chen_2008, Nair_2008, Mak_2008}, and new device concepts have been proposed to make use of those special features \cite{Katsnelson_2006, Cheianov_2007}. Recently, a new type of terahertz laser based on graphene has been proposed \cite{Ryzhii_2007, Ryzhii_2010}. Analytical model calculations considering relaxation involving optical phonon emission have shown that with sufficiently strong optical pumping, population inversion and negative ac conductivity may be achieved in the terahertz frequency range \cite{Ryzhii_2007}. This results from the fast optical phonon relaxation followed by an acoustic phonon relaxation bottleneck for excited electrons and holes with energies $E=E_F\pm\hbar\omega_{\rm{o}}$, where $E_F$ is the Fermi level, and $\hbar\omega_{\rm{o}}$ is the optical phonon energy. Some initial experimental results also have been reported suggesting stimulated terahertz emissions in graphene \cite{Karasawa_2010}. The threshold pumping power needed to obtain population inversion depends on the carrier recombination mechanism \cite{Ryzhii_2007}. Previous studies on quantum-well systems \cite{Goodnick_PRB_1988} have shown that electron-electron (e-e) scattering is responsible for the fast relaxation of photo-excited carriers, and recent experiments on graphene \cite{Newson_2009, Choi_APL_2009} also report ultrafast carrier kinetics. Therefore, it is essential to treat carrier dynamics rigorously to simulate the behavior of photo-excited carriers in graphene.

Here we solve a time-dependent Boltzmann transport equation (BTE) for two-dimensional (2D) graphene with optical pumping. We consider carrier scattering and recombination-generation (R-G) mechanisms including acoustic and optical phonon scattering \cite{Lazzeri_2005, Ando_2006, [{}][{ and references therein.}]Perebeinos_2010}, charged impurity scattering \cite{Hwang_PRB_2008}, photo-generation \cite{Nair_2008}, spontaneous optical transition \cite{Vasko_2008}, and e-e scattering that includes impact ionization and Auger recombination \cite{Rana_2007}. Using the numerical solutions of the BTE, we study the influence of each scattering mechanism on the properties of optically pumped graphene, compare the results with those from previous approaches with phonon scattering only \cite{Ryzhii_2007}, and address the challenges involved in realizing population inversion. We solve the BTE as
\begin{equation}
\frac{f_{s\bf{k}}^{n+1}-f_{s\bf{k}}^{n}}{\Delta t}=\frac{q\mathscr{E}_x}{\hbar}\frac{\partial f_{s\bf{k}}^{n}}{\partial k_x}+\hat{C}f_{s\bf{k}}^{n}+G_{s\bf{k}}^{n},\label{eq1}
\end{equation}
where $f_{s\bf{k}}^{n}$ is the distribution function at the n{\it th} time step for a state with 2D wavevector ${\bf{k}}=(k_x,k_y)$ in the conduction ($s=+1$) or in the valence ($s=-1$) band, {\it q} is the unit charge, $\hbar$ is the reduced Planck constant, $\mathscr{E}_x$ is the electric field along the {\it x}-direction, $\hat{C}f_{s\bf{k}}^{n}$ and $G_{s\bf{k}}^{n}$ are the collision integral and the R-G rates, and $\Delta t$ is the time interval, which is 0.05 $\sim$ 10 fs in our simulation. We treat a bulk graphene under uniform illumination, so $f_s$ has no spatial dependence.

The collision integral treats carrier scattering as
\begin{eqnarray*}
\hat{C}f_{s\bf{k}}&=&\sum_{\bf{k'}}f_{s'\bf{k'}}(1-f_{s\bf{k}})S_{\bf{k'},\bf{k}}-\sum_{\bf{k'}}\:f_{s\bf{k}}(1-f_{s'\bf{k'}})S_{\bf{k},\bf{k'}},
\end{eqnarray*}
where $S_{\bf{k},\bf{k'}}$ is the transition rate from $\bf{k}$ to $\bf{k'}$, and $ss'=\pm 1$ means the intra- and inter-band transition respectively. For the two acoustic phonon modes, $\Gamma_{\rm{LA}}$ and $\Gamma_{\rm{TA}}$, the sum of the two contributions can be treated as isotropic using an averaged sound velocity $\upsilon_S$ as
\begin{eqnarray*}
S_{\bf{k},\bf{k'}}&=&\frac{2\pi}{\hbar}\frac{D_A^2\beta}{2A\upsilon_F\rho_m\upsilon_S}(N_{\omega}+\frac{1}{2}\mp\frac{1}{2})\\
&&{\times}\:\delta_{\bf{k'},\bf{k}\pm\bm{\beta}}\delta(k'-k\mp{\beta\upsilon_S}/{\upsilon_F}),
\end{eqnarray*}
where $D_A$ is the acoustic phonon deformation potential, $\bm{\beta}$ is the 2D phonon wavevector with magnitude $\beta$, $\rho_m$ is the mass density, {\it A} is the area of the sample, $\upsilon_F$ is the Fermi velocity, $N_{\omega}$ is the Bose-Einstein phonon occupation number, $\delta$ is the Dirac delta function, and $\mp$ (or $\pm$) signs represent the absorption and emission of phonons. For $D_A$ = 7.1 eV and $\upsilon_S$ = 17.3 $\rm{km/s}$ \cite{Perebeinos_2010}, we obtain a low field mobility of $\sim190,000$ $\rm{cm^2V^{-1}s^{-1}}$ for carrier density of $\sim10^{12}$ $\rm{cm^{-2}}$ as in Ref. \cite{Chen_2008}. Acoustic phonon scattering is treated to be inelastic for a rigorous treatment of energy relaxation and broadening of distribution functions.
Electron-phonon (e-ph) coupling of long-wavelengh optical phonon modes, $\Gamma_{\rm{LO}}$ and $\Gamma_{\rm{TO}}$, is expressed as \cite{Lazzeri_2005, Ando_2006}
\begin{equation}
S_{\bf{k},\bf{k'}}=\frac{2\pi}{\hbar}\frac{D_{\rm{o}}^2}{2A\upsilon_F\rho_m\omega_{\rm{o}}}(N_{\omega}+\frac{1}{2}\mp\frac{1}{2})\delta(k'-k\mp\frac{\omega_{\rm{o}}}{\upsilon_F}),\label{eq2}
\end{equation}
where $D_{\rm{o}}=3\sqrt{2}J_1/2\sim$ 11 eV$\mathrm{\mathring{A}^{-1}}$ is the optical phonon deformation potential with a coupling constant of $J_1$ = 5.3 eV$\mathrm{\mathring{A}^{-1}}$ \cite{Perebeinos_2010}, and $\hbar\omega_{\rm{o}}$ = 197 meV. For zone edge phonon modes, only the transverse mode $\rm{K_{TO}}$ contributes to the carrier scattering, and the transition rate is given by Eq. (\ref{eq2}) multiplied by $({1-ss'\cos(\theta_{\bf{k}}}-\theta_{\bf{k'}}))/2$ \cite{Lazzeri_2005}, with $D_{\mathrm{o}}=3J_1\sim16$ eV$\mathrm{\mathring{A}^{-1}}$, $\hbar\omega_{\mathrm{o}}=157$ meV, and $\theta_{\bf{k}}$ is the angle of $\bf{k}$. 

While the charged impurity scattering \cite{Hwang_PRB_2008} degrades the graphene mobility significantly due to momentum relaxation, we find that it has little effect on the optical properties (the main focus of this work) because it does not relax energy.

Coulomb scattering among electrons contributes to the R-G process through impact ionization and Auger recombination \cite{Rana_2007}, and it can also thermalize distribution functions without changing the total carrier density. For Auger recombination and impact ionization, there are four processes involved, CCCV, VVCV, CVCC, and CVVV, where C and V represent the conduction and valence bands, and the second/third and first/fourth letters represent the initial and final states of the two electrons involved. For example, the transition rate from $\bf{k}$ to $\bf{k'}$ in the conduction band due to CVCC is expressed as
\begin{eqnarray*}
S_{\bf{k},\bf{k'}}&=&\frac{2\pi}{\hbar^2\upsilon_F}\sum_{\bf{k_1}}|M|^2\delta(|\bf{k'}|+|\bf{k_1}+\bf{k}-\bf{k'}|-|\bf{k}|+|\bf{k_1}|)\\
&&{\times}f_{-1,\bf{k_1}}(1-f_{+1,\bf{k_1}+\bf{k}-\bf{k'}}),
\end{eqnarray*}
where $|M|^2 = |M_d|^2+|M_e|^2-|M_d-M_e|^2=2|M_d|^2+2|M_e|^2-M_d^*M_e-M_dM_e^*$, and $M_d$ and $M_e$ are the direct and exchange matrix elements. To treat the K-K' valley degeneracy, we double the $2|M_d|^2$ and $2|M_e|^2$ terms. Coulomb interactions are calculated using the random phase approximation (RPA) \cite{Hwang_PRB_2008}. The resulting collision integral in Eq. (\ref{eq1}) becomes four-dimensional, and it can be reduced to 2D using the momentum and energy conservation conditions \cite{Rana_2007}. Processes conserving the number of electron-hole pairs, i.e., CCCC, VVVV, CCVV, and CVCV, are treated in a similar way.

$G_{s\bf{k}}$ in Eq. (\ref{eq1}) includes photo-generation and spontaneous emission. The photo-generation rate is
\begin{equation}
G_{s\bf{k},{\it opt}}=\alpha\frac{\pi^2\upsilon_FP_{opt}}{\hbar\omega_{opt}^2}(f_{s'\bf{k}}-f_{s\bf{k}})\delta(k-\frac{\omega_{opt}}{2\upsilon_F}),\label{eq3}
\end{equation}
where $\alpha$ is the fine structure constant, $\hbar\omega_{opt}\equiv E_{opt}$ is the incident photon energy, $P_{opt}$ is the pumping power, and $ss'=-1$. From Eq. (\ref{eq3}), we can show that the graphene with $f_{+1}=0$ and $f_{-1}=1$ has the absorbance of $\pi\alpha\sim$ 2.3 $\%$ \cite{Nair_2008}. The spontaneous emission is treated using the approach in Ref. \cite{Vasko_2008}, but its contribution is much smaller than those from other processes.

Using the models above, we numerically solve the BTE on a 2D {\it k}-grid. We first simulate optically pumped graphene without considering e-e scattering, where e-ph scattering is the main mechanism for energy relaxation as in the previous approaches \cite{Ryzhii_2007, Ryzhii_2010}. Then we introduce e-e scattering and see how the results change. We also explore the effects of $E_{opt}$, $P_{opt}$, and $E_F$ on the properties of optically pumped graphene. All simulation results are for 300 K, zero $\mathscr{E}_x$, and no charged impurity. Photo-current with finite $\mathscr{E}_x$ due to the transport of photo-excited carriers will be discussed elsewhere.

\begin{figure}
\includegraphics[scale=0.45]{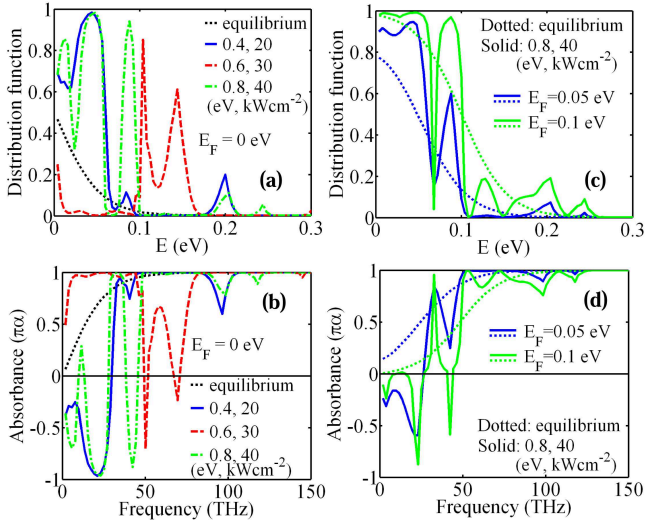}
\caption {\label{fig1}Simulation results for optically pumped graphene under steady-state with e-ph scattering only. (a) $f_{+1}$ vs. {\it E} and (b) absorbance for $E_F$ = 0 eV, $E_{opt}$ = 0.4/0.6/0.8 eV, and $P_{opt}$ = 20/30/40 $\mathrm{kWcm^{-2}}$. (c) $f_{+1}$ vs. {\it E} and (d) absorbance for $E_F$ = 0.05/0.1 eV, $E_{opt}$ = 0.8 eV, and $P_{opt}$ = 40 $\mathrm{kWcm^{-2}}$. Population inversion is achieved and modulated by $E_{opt}$.}
\end{figure}

Figure \ref{fig1} shows simulation results under steady-state low power optical pumping considering e-ph scattering only. In all following results, $f_{\pm1}$ means $f_{\pm1,\bf{k}}$ averaged over $\bf{k}$ with $|{\bf{k}}|=k$ and $E=\pm\hbar\upsilon_Fk$. In Fig. \ref{fig1}(a) for $E_F$ = 0 eV, $f_{+1}$ under optical pumping dramatically deviates from equilibrium with peaks modulated by $E_{opt}$. For $E_{opt}=0.6$ eV, for example, photo-excited electrons first generated at {\it E} = 0.3 eV pile up at {\it E} $\sim$ 0.1 eV and 0.15 eV due to the emission of $\Gamma_{\mathrm{LO/TO}}$ and $\mathrm{K_{TO}}$ phonons. For $E_F$ = 0 eV, photo-excited holes pile up symmetrically in the valence band. Due to interband optical phonon scattering, the piled-up electrons and holes recombine with holes and electrons in the opposite bands, so the carriers with $E<0.1$ eV are depleted in Fig. \ref{fig1}(a). Figure \ref{fig1}(b) shows the optical absorbance due to interband transition, $\pi\alpha(f_{-1}-f_{+1})$, vs. frequency $2|E|/h$, where {\it h} is the Planck constant. The absorbance becomes negative, i.e., population inversion is achieved, at specific frequency ranges depending on $E_{opt}$. Simulation results for $E_F=0.05$ eV and 0.1 eV are shown in Figs. \ref{fig1}(c)-(d). For finite $E_F$'s, $f_{+1}$ and $f_{-1}$ are not symmetric, and the frequency and the height of the population inversion peaks depend on $E_F$. In general, as $E_F$ increases, a smaller energy region becomes available for the minority carriers to experience an acoustic phonon bottleneck, so a larger pumping power is required to invert the population.

\begin{figure}
\includegraphics[scale=0.45]{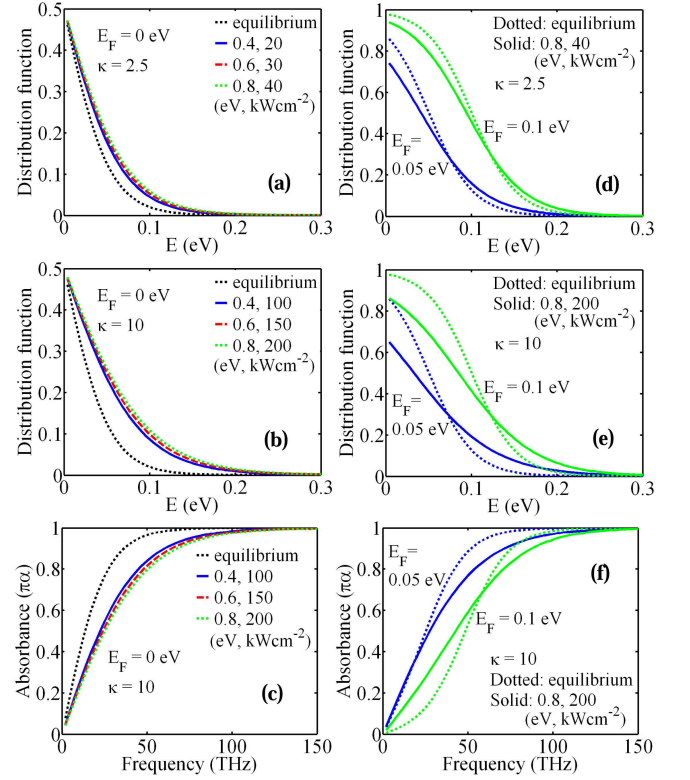}
\caption{\label{fig2}Simulation results for optically pumped graphene under steady-state with e-e scattering considered. (a) $f_{+1}$ vs. {\it E} for $E_F$ = 0 eV, $\kappa$ = 2.5, $E_{opt}$ = 0.4/0.6/0.8 eV, and $P_{opt}$ = 20/30/40 $\mathrm{kWcm^{-2}}$. (b) $f_{+1}$ vs. {\it E} and (c) absorbance with increased pumping powers and $\kappa$ = 10. (d) $f_{+1}$ vs. {\it E} for $E_F$ = 0.05/0.1 eV, $\kappa$ = 2.5, $E_{opt}$ = 0.8 eV, and $P_{opt}$ = 40 $\mathrm{kWcm^{-2}}$. (e) $f_{+1}$ vs. {\it E} and (f) absorbance with increased pumping powers and $\kappa$ = 10. Distribution functions are thermalized with elevated $T_e$'s, and population inversion is not achieved.}
\end{figure}

Figure \ref{fig2} shows simulation results for optically pumped graphene under steady-state with e-e scattering considered. In Fig. \ref{fig2}(a) for $E_F$ = 0 eV and dielectric constant $\kappa$ = 2.5, $f_{+1}$ with optical pumping does not deviate much from equilibrium, and there is little $E_{opt}$-dependence unlike in the case of Fig. \ref{fig1}(a). We find that under photo-excitation the distribution function remains equilibrium with an elevated electronic temperature $T_e$ determined from the Fermi-Dirac distribution. In Fig. \ref{fig2}(a), $T_e$ = 381 K, 402 K, and 419 K for $E_{opt}$ = 0.4 eV, 0.6 eV, and 0.8 eV respectively. In Fig. \ref{fig2}(b), although we increase $P_{opt}$ by five times and use a high $\kappa$ = 10 to reduce the Coulomb interaction, the distribution functions are more broadened with higher $T_e$'s (497 K, 529 K, and 553 K) and still far from population inversion, and the absorbance never goes below zero in Fig. \ref{fig2}(c). Here we should note that in RPA \cite{Hwang_PRB_2008}, the Coulomb interaction does not scale as $\kappa^{-2}$ as expected from the unscreened model. For $E_F$ = 0 eV, for example, the RPA coincides with the unscreened model with $\kappa+{q^2}/({8\hbar\upsilon_F\varepsilon_0})\sim\kappa+3.5$ where $\varepsilon_0$ is the vacuum permittivity, so changing $\kappa$ from 2.5 to 10 results in the Coulomb interaction being reduced by about five times, not by 16 times. Simulation results for $E_F$ = 0.05 eV and 0.1 eV are shown in Figs. \ref{fig2}(d)-(f). In Fig. \ref{fig2}(d), population inversion is never achieved for $\kappa$ = 2.5 and the same $P_{opt}$'s as those in Fig. \ref{fig1}(c), and the distribution functions are thermalized with $T_e$ = 408 K and 379 K for $E_F$ = 0.05 eV and 0.1 eV respectively. With $\kappa$ = 10 and larger pumping powers in Fig. \ref{fig2}(e)-(f), distribution functions are more broadened with elevated $T_e$'s (536 K and 494 K) instead of showing multiple peaks as in Fig. \ref{fig1}(c).

\begin{figure}
\includegraphics[scale=0.43]{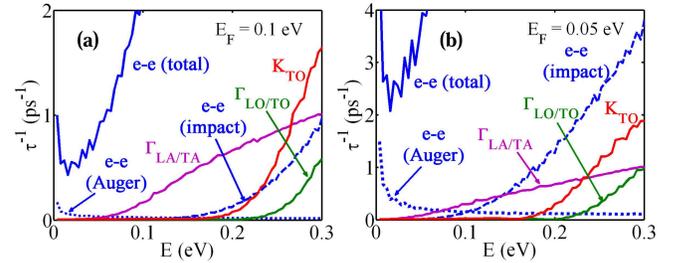}
\caption{\label{fig3}Simulation results for $\tau^{-1}$ vs. {\it E} (conduction band, $\kappa$ = 10) under equilibrium with (a) $E_F$ = 0.1 eV and (b) $E_F$ = 0.05 eV. Coulomb scattering rates are much larger than e-ph scattering rates and increase with decreasing $E_F$.}
\end{figure}

Therefore, e-e scattering significantly broadens the distribution functions washing out peaks due to cascade emission of optical phonons \cite{Goodnick_PRB_1988}. To observe this quantitatively, we calculate the carrier lifetime $\tau_{s,\bf{k}}=\sum_{\bf{k'}}(1-f_{s'\bf{k'}})S_{\bf{k},\bf{k'}}$, which is the characteristic time it takes for a carrier in the state with $\bf{k}$ to scatter out to other states. Figure \ref{fig3} shows results for $\tau^{-1}$ vs. {\it E} for conduction band under equilibrium with $E_F$ = 0.1 eV and 0.05 eV and $\kappa$ = 10, where $\tau$ is the average of $\tau_{\bf{k}}$ for $|{\bf{k}}|=k$. The contribution from e-e scattering scales linearly with {\it E}, and it is much larger than those from e-ph scattering and increases with decreasing $E_F$ due to reduced screening. 

\begin{figure}
\includegraphics[scale=0.43]{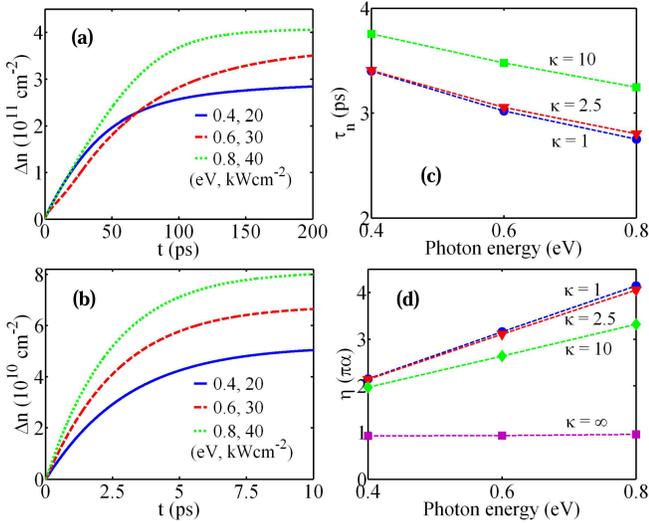}
\caption{\label{fig4}Simulation results for $\Delta n$ vs. {\it t} with optical pumping ($E_{opt}$ = 0.4/0.6/0.8 eV, $P_{opt}$ = 20/30/40 $\mathrm{kWcm^{-2}}$) for $E_F$ = 0 eV with (a) e-ph scattering only and (b) e-e scattering considered with $\kappa$ = 2.5. (c) Relaxation time vs. $E_{opt}$ and (d) absorption efficiency vs. $E_{opt}$ for various $\kappa$'s for the same optical pumping conditions.}
\end{figure}

So far we have mainly discussed steady-state results, but transient responses \cite{Hertel_2000} are also important in understanding carrier dynamics in graphene. In Figs. \ref{fig4}(a)-(b), we compare the photocarrier density $\Delta n=n-n_0$ vs. {\it t} for $E_F$ = 0 eV with and without e-e scattering ($\kappa$ = 2.5), where {\it n} is the total electron density, $n_0$ is the equilibrium electron density, and the optical pumping starts at time {\it t} = 0 ps. With e-e scattering, steady-state is quickly achieved, and $\Delta n$, which in turn determines the photo-current response \cite{Avouris_2010}, is much smaller than with e-ph scattering only. We can fit the $\Delta n$ vs. {\it t} curves using $\Delta n$ = $A_n\tau_n(1-\mathrm{exp}(-t/\tau_n))$ and extract the relaxation time $\tau_n$ and the absorption efficiency $\eta=A_n/R_{opt}$, where $R_{opt}=P_{opt}/E_{opt}$. For e-ph scattering only in Fig. \ref{fig4}(a), $\tau_n$ is very large (45 $\sim$ 60 ps) and increases with $E_{opt}$ while it is much shorter for e-e scattering ($\sim$ 3 ps) in Fig. \ref{fig4}(b). As shown in Fig. \ref{fig4}(c), $\tau_n$ increases with increasing $\kappa$ and somewhat decreases with increasing $E_{opt}$. Figure \ref{fig4}(d) shows the $\eta$ vs. $E_{opt}$ results. For $\kappa=\infty$ (i.e. e-ph scattering only), $\eta$ is close to $\pi\alpha$ and independent of $E_{opt}$. For finite $\kappa$'s, however, $\eta$ is greater than $\pi\alpha$, which indicates carrier multiplication. The $\eta$ increases with decreasing $\kappa$ due to the increased Coulomb interaction, and it increases with increasing $E_{opt}$ because high energy photocarriers have higher ability to initiate impact ionization. A recent theoretical study has reported that with pulse excitations, impact ionization leads to dramatically increased photocarrier densities \cite{Winzer_2010}. However, for continuous excitation (the main focus of this work), e-e scattering leads to a lower steady-state $\Delta n=A_n\tau_n$ because Auger recombination becomes stronger and recombination due to optical phonon emission is boosted by the broadened distribution functions, which all result in a reduced $\tau_n$. We find that $T_e$ follows the similar time evolution as $\Delta n$, and the steady-state $T_e$ under optical pumping depends on $E_{opt}$, $P_{opt}$, and $\kappa$ as shown in Fig. \ref{fig2}. We also find that the decay time constant, after the optical pumping is switched off, is slightly different than the rise time constant due to the difference in the scattering rate determined by the distribution function in the photo-excited state.

In summary, we numerically solved a time-dependent BTE for bulk graphene considering photo-generation and relevant carrier scattering and R-G mechanisms to explore properties of optically pumped graphene and possibilities to achieve population inversion. Simulation results showed that when carrier energies are relaxed by e-ph scattering only, it is possible to achieve population inversion at frequencies controlled by incident photon energy and graphene Fermi level. A more realistic model considering Coulomb scattering, however, showed that distribution functions are significantly broadened, and it becomes hard to achieve population inversion even with much higher pumping power. The results stress the importance of e-e scattering in carrier dynamics in graphene and suggest that the Coulomb interaction among carriers must be significantly suppressed, such as by bandgap opening in bilayer graphene, to realize graphene-based terahertz lasers. However, e-e scattering leads to significant carrier multiplication.

We thank Prof. Mark S. Lundstrom at Purdue University for helpful discussions. Computational support was provided by the Network for Computational Nanotechnology, supported by the National Science Foundation.

\end{document}